\documentclass[%
 aip,
 rsi,%
 amsmath,amssymb,
 reprint,%
]{revtex4-1}

\usepackage{graphicx}
\usepackage{dcolumn}
\usepackage{bm}
\usepackage{textgreek} 
\usepackage{xcolor} 

\begin{document}

\title{Integrated impedance bridge for absolute capacitance measurements at cryogenic temperatures and finite magnetic fields}

\author{G.J.~Verbiest}
\affiliation{JARA-FIT and 2nd Institute of Physics, RWTH Aachen University, 52056 Aachen, Germany, EU}
\affiliation{Current address: 3mE, Delft University of Technology, 2628 CD Delft, The Netherlands, EU}

\author{H.~Janssen}
\affiliation{JARA-FIT and 2nd Institute of Physics, RWTH Aachen University, 52056 Aachen, Germany, EU}

\author{D.~Xu}
\affiliation{JARA-FIT and 2nd Institute of Physics, RWTH Aachen University, 52056 Aachen, Germany, EU}
\affiliation{QuTech and Kavli Institute of Nanoscience, Delft University of Technology, 2600 GA Delft, The Netherlands, EU.}

\author{X.~Ge}
\affiliation{JARA-FIT and 2nd Institute of Physics, RWTH Aachen University, 52056 Aachen, Germany, EU}

\author{M.~Goldsche}
\author{J.~Sonntag}
\author{T.~Khodkov}
\author{L.~Banszerus}
\affiliation{JARA-FIT and 2nd Institute of Physics, RWTH Aachen University, 52056 Aachen, Germany, EU}
\affiliation{Peter Gr{\"u}nberg Institute (PGI-8/9), Forschungszentrum J{\"u}lich, 52425 J{\"u}lich, Germany, EU}

\author{N.~von den Driesch}
\author{D.~Buca}
\affiliation{Peter Gr{\"u}nberg Institute (PGI-8/9), Forschungszentrum J{\"u}lich, 52425 J{\"u}lich, Germany, EU}

\author{K.~Watanabe}
\author{T.~Taniguchi}
\affiliation{National Institute for Materials Science, Tsukuba, Ibaraki 305-0047, Japan}

\author{C.~Stampfer}
\affiliation{JARA-FIT and 2nd Institute of Physics, RWTH Aachen University, 52056 Aachen, Germany, EU}
\affiliation{Peter Gr{\"u}nberg Institute (PGI-8/9), Forschungszentrum J{\"u}lich, 52425 J{\"u}lich, Germany, EU}

\begin{abstract}
We developed an impedance bridge that operates at cryogenic temperatures (down to 60 mK) and in perpendicular magnetic fields up to at least 12 T.
This is achieved by mounting a GaAs HEMT amplifier perpendicular to a printed circuit board containing the device under test and thereby parallel to the magnetic field.
The measured amplitude and phase of the output signal allows for the separation of the total impedance into an absolute capacitance and a resistance.
Through a detailed noise characterization, we find that the best resolution is obtained when operating the HEMT amplifier at the highest gain.
We obtained a resolution in the absolute capacitance of 6.4~aF$/\sqrt{\textrm{Hz}}$ at 77 K on a comb-drive actuator, while maintaining a small excitation amplitude of 15~$k_\text{B} T/e$.
We show the magnetic field functionality of our impedance bridge by measuring the quantum Hall plateaus of a top-gated hBN/graphene/hBN heterostructure at 60~mK with a probe signal of 12.8~$k_\text{B} T/e$.
\end{abstract}

\maketitle

\section{Introduction\label{intro}}

Electronic and electromechanical devices are characterized by an impedance that defines their functionality.
Thus, an accurate measurement system for impedances is thus of crucial importance to develop, characterize, and optimize electronic and electromechanical devices. \cite{roadmap2015}
Impedances are commonly measured with bridge circuits or LCR-meters.
The ongoing miniaturization of electronic and electromechanical devices push these measurement techniques to the limit.
For example, the density of available electronic states in nanostructures becomes finite which results in a so-called quantum capacitance \cite{luryi1988,john2004} or the displacement of electromechanical devices enter the nanoscale. \cite{goldsche2018raman,goldsche2018fab}
Both these examples result typically in capacitance changes of only a few aF.
Standard measurement techniques cannot resolve these changes due to parasitic impedances arising from lengthy cables connecting the device under test (DUT) and the measurement system.

Direct measurements of the density of states via the quantum capacitance have been successful for among others, graphene \cite{xia2009,droscher2010,li2011,hunt2013,chen2013,yu2013,kliros2015,zibrov2017,cao2018} and carbon nanotube \cite{ilani2006} devices as well as GaAs-based devices containing a two-dimensional (2D) electron gas. \cite{smith1985,stern1983}
In such systems with a low density of states the total capacitance needs to be expressed by the quantum capacitance $C_\text{q}$ in series to the geometric capacitance $C_\text{g}$
leading to a total capacitance $C_\text{tot} = C_\text{q}C_\text{g}/(C_\text{q}+C_\text{g})$, which is also strongly depending on the density of states.
Measurements of this quantity were performed with a bridge circuit, \cite{agilent2009,foote1986,andeen1988,gokirmak2009,ashoori1992} a LCR meter, or even a scanning tunneling microscope. \cite{martin2008}
The latter provides local information whereas the former characterizes the entire device.
In order to fully resolve fine features in their electronic bandstructure such as van Hove singularities in carbon nanotubes, a probe signal on the order of the characteristic thermal energy $k_\text{B} T/e$ is required. \cite{kliros2015}
However, bridge circuits and LCR meters usually include some lengthy cables which give rise to a parasitic capacitance in the order of hundreds of picofarads.
This large parasitic capacitance results in a huge attenuation of the probe signal which, in combination with the required $k_\text{B} T/e$ excitation, pushes attofarad capacitance changes below the resolution limit.
To reduce the effect of the parasitic capacitance of the cables, Hazeghi {\it et al.} used a high-electron mobility transistor (HEMT) as an impedance-matching amplifier in an integrated capacitance bridge. \cite{hazeghi2011,li2011,hunt2013,cao2018}
A resolution of 60~aF$\mathrm{/\sqrt{Hz}}$ at 300~K and 21~aF$\mathrm{/\sqrt{Hz}}$ at 4.2~K was achieved on a top-gated graphene device with probe signals on the order of $k_\text{B} T/e$.
However, a problem arises if the density of states approaches zero.
As $C_q$ approaches zero, also the {\it conductance} approaches zero due to the absence of charge carriers.
Depending on the frequency of the probe signal, the measured signal becomes a function of both and thus depends on the total impedance.

\begin{figure}[!t]
  \centering
  \includegraphics[width=70mm]{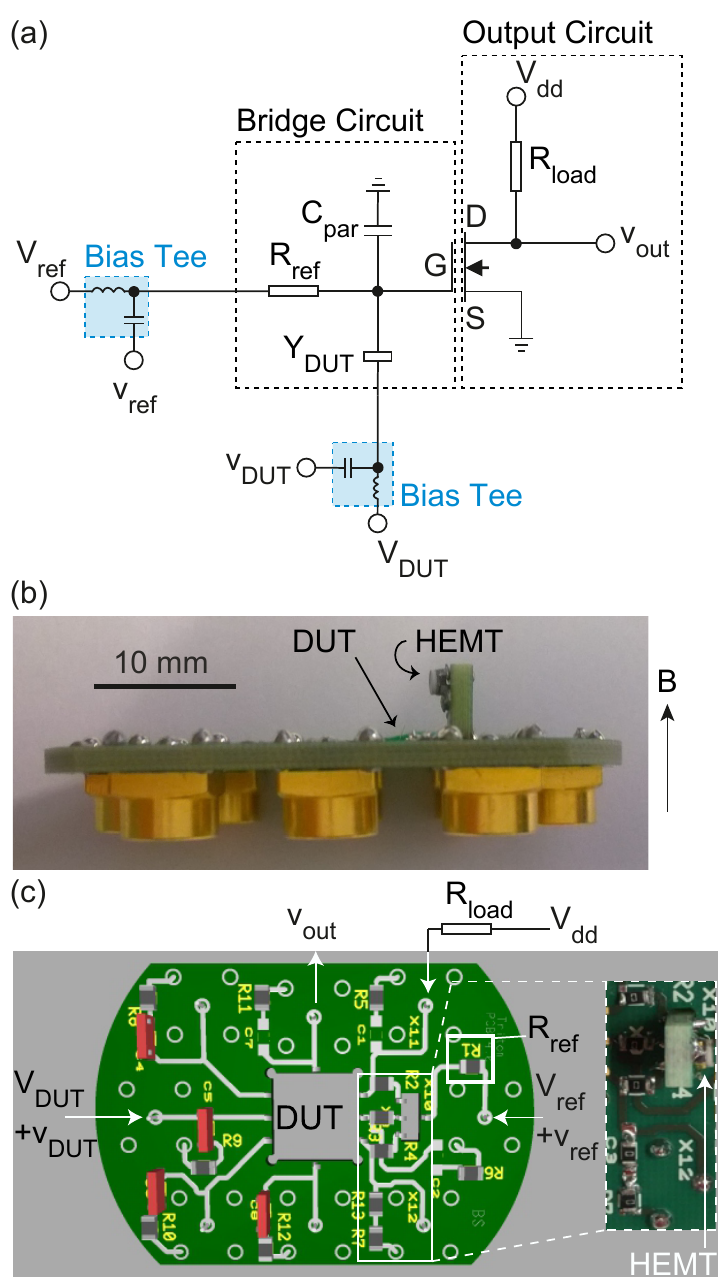}
  \caption{(a) Schematic overview of the measurement circuit, which consists of a bridge circuit and an output circuit that are coupled via a HEMT amplifier. The bridge circuit contains a reference resistor of 1 M$\mathrm{\Omega}$, a parasitic capacitance $C_\text{par}$ to ground, and an unknown admittance $Y_\text{DUT}$. The output circuit has a load resistor of 1 k$\mathrm{\Omega}$ and an output line for $V_\text{out}$. The bias tees used for summing the AC and DC signal (blue) are not on the PCB. (b) A side view (optical image) of the PCB shows the perpendicular placement of the HEMT on the PCB, which allows for measurements at high magnetic fields. (c) top view (optical image) of the PCB including the placement of the different components and the electrical connections.}
  \label{fig1}
\end{figure}

In this work, we expand the previously reported integrated capacitance bridge \cite{hazeghi2011} into an integrated impedance bridge to determine not only the absolute capacitance but also the resistance.
We achieve this by performing a thorough analysis, including noise optimization, of both the amplitude and the phase of the measured signal.
Our work differs at four additional points from existing literature: (i) we place the HEMT perpendicular on a printed circuit board (PCB) to allow for magnetic field dependent measurements, (ii) we use a much smaller reference resistor to operate the bridge at higher frequencies, (iii) we show that the impedance bridge operates at temperatures down to 60~mK, and (iv) we show through a detailed noise characterization that the best impedance resolution is obtained when operating the HEMT at its highest amplification.
Ultimately, we achieve a resolution of 6.4~aF$\mathrm{/\sqrt{Hz}}$ at a temperature of 77~K.
To show the magnetic field applicability of our circuit, we measure the quantum Hall plateaus emerging in a top-gated hBN/graphene/hBN heterostructure in magnetic fields up to 12~T at 60~mK.

\section{Bridge design and operation\label{sect1}}

The bridge circuit (see Fig.~\ref{fig1}a) consists of a reference resistor and an impedance-matching amplifier to eliminate the effect of large parasitic cable capacitances.
To ensure functionality of the bridge down to cryogenic temperatures, we use a (GaAs-based) HEMT as impedance-matching amplifier. \cite{hazeghi2011}
The HEMT in our bridge is a packaged FHX35LG transistor \cite{fujitsu} with a gate capacitance of $\sim$0.5 pF that is part of a small remaining parasitic capacitance $C_\text{par}$.
A gate bias $V_\text{ref}$ of -1~V fully depletes the 2D electron gas in the GaAs-based HEMT.

The reference resistor $R_\text{ref}$ is used to balance the signal across the unknown impedance $Z_\text{DUT} = Y_\text{DUT}^{-1}$, where $Y_\text{DUT}$ is the corresponding admittance of the device under test (DUT).
As the HEMT gate (G) is biased through $R_\text{ref}$, $R_\text{ref}$ should be much smaller than the HEMT DC gate resistance across all temperatures.
In addition, $R_\text{ref}$ forms together with $C_\text{par}$ and the capacitive contribution of $Z_\text{DUT}$, a low-pass filter.
We choose $R_\text{ref}$ equal to the impedance of the expected $Z_\text{DUT}$ parallel to a parasitic capacitance $C_\text{par}$ at the measurement frequency to prevent additional shunting of the signal across $Z_\text{DUT}$.
To satisfy these constraints and to maximize the bandwidth, we use a reference resistance $R_\text{ref}$ of 1 M$\mathrm{\Omega}$ with a low temperature coefficient (SMD type thick film resistor).

The output circuit (see Fig.~\ref{fig1}a) consists of a load resistor $R_\text{load}$ and the drain (D) to source (S) resistance of the HEMT.
$R_\text{load}$ is used to bias the HEMT drain with the voltage $V_\text{dd}$.
To ensure stable operation, $R_\text{load}$ must be larger than the resistance of the cables.
Therefore, we choose a $R_\text{load}$~=~1~k$\mathrm{\Omega}$.
The maximum amplification is expected when the drain-source resistance of the HEMT is roughly equal to $R_\text{load}$.
We always place $R_\text{load}$ at room temperature to minimize the heat load of the output circuit.
While measuring in a dilution refrigerator (a Triton 200 system), the temperature, as measured with a built-in calibrated RuO$_\text{2}$ sensor, increased from 15~mK to 50~mK when biasing the HEMT drain with $V_\text{dd} = 0.55$~V.

\begin{figure}[!t]
	\centering
	\includegraphics[width=85mm]{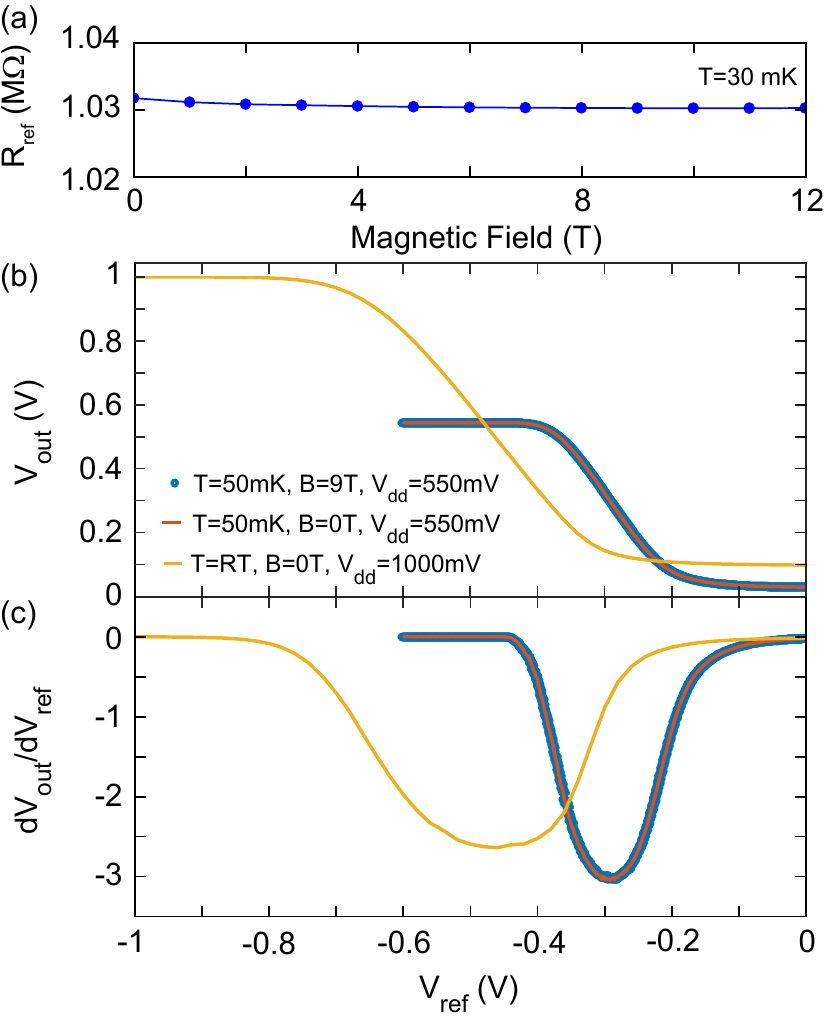}
	\caption{(a) The measured reference resistance $R_\text{ref}$ as a function of magnetic field at 30 mK. Comparing the value at $B = 12$~T with the one at 0~T
    indicates a negligible change of 0.1\%. (b) DC output signal $V_\text{out}$ as a function of the applied DC potential $V_\text{ref}$ on the gate
    of the HEMT and (c) the derivative of (b) with respect to $V_\text{ref}$, which defines the gain of the HEMT. We observe a change when going from room temperature (RT) to 50 mK, but a negligible difference between the curves at 0~T and 9~T, which are both taken at 50 mK.}
	\label{fig3}
\end{figure}

For successful operation of the circuit in a perpendicular magnetic field, we place the HEMT on a small PCB that is mounted perpendicular to the PCB containing $R_\text{ref}$ and $Z_\text{DUT}$ (see Fig.~\ref{fig1}(b) and (c)).
The electrical connections to the PCB are made via SMP thru hole PCB mounts.
We characterized both $R_\text{ref}$ and the HEMT in a Triton 200 system at mK temperatures for different applied magnetic fields.
$R_\text{ref}$ is approximately 0.98 M$\mathrm{\Omega}$ at room temperature and 1.03 M$\mathrm{\Omega}$ at 30 mK.
Both values are within the 3\% tolerance given by the manufacturer.
As Fig. \ref{fig3}a shows, $R_\text{ref}$ only changes by 0.1\% when applying a magnetic field of 12~T perpendicular to the PCB containing $R_\text{ref}$.
Similarly, the operation of the HEMT remains unaffected when a magnetic field is applied parallel to the drain-source channel of the HEMT (see Figs.~\ref{fig3}b and \ref{fig3}c).
We conclude that the resulting Lorentz forces acting on the 2D electron gas have no effect and thus that the integrated impedance bridge remains fully functional.

In operation, two AC signals at the same frequency are used, of which one is sent through a reference resistance $R_\text{ref}$ ($v_\text{ref}$) and one through the DUT ($v_\text{DUT}$).
The amplitude and phase of these signals are set such that they interfere destructively at the so-called bridge point at the gate of the HEMT.
The signal $v_\text{b}$ at the bridge point is an average of the applied AC signals $v_\text{ref}$ and $v_\text{DUT}$,

\begin{equation}
v_\text{b} = \frac{Y_\text{DUT} \cdot v_\text{DUT} + Y_\text{ref} \cdot v_\text{ref}}{Y_\text{ref} + Y_\text{DUT} + Y_\text{par}},
\label{eq1}
\end{equation}

\noindent
where $Y_\text{ref}$, $Y_\text{DUT}$, and $Y_\text{par}$ are the admittances of $R_\text{ref}$, $Z_\text{DUT}$, and $C_\text{par}$, respectively.
When balanced ($v_\text{b} = 0$~V), the amplitude and phase of $Y_\text{DUT}$ are given by

\begin{equation}
Y_\text{DUT} = -\frac{Y_\text{REF} \cdot v_\text{REF}}{v_\text{DUT}}.
\label{eq2}
\end{equation}

\noindent
As $Y_\text{DUT}$ is the only quantity that is tunable with the DC voltage $V_\text{DUT}$, we need to characterize the other admittances in the circuit only once.
The signal $v_\text{b}$ is amplified $A_\text{HEMT}$ times by the HEMT into the output voltage $v_\text{out} = A_\text{HEMT} \cdot v_\text{b}$, which is measured with a lock-in amplifier.
The output voltage $v_\text{out}$ is in all measurements below 0.5 mV, which allows us to neglect any nonlinear contribution from the HEMT transduction to $A_\text{HEMT}$.
The measured amplitude and phase of the output voltage $v_\text{out}$ is used in Eq.~(\ref{eq1}) to compute the unknown $Y_\text{DUT}$.
The sensitivity $S$ of the circuit to a change in $Y_\text{DUT}$ is given by the derivative of Eq.~(\ref{eq1}) with respect to $Y_\text{DUT}$

\begin{equation}
S = \frac{Y_\text{ref} \cdot \left(v_\text{DUT} - v_\text{ref}\right) + Y_\text{par} \cdot v_\text{DUT}}{\left(Y_\text{ref} + Y_\text{DUT} + Y_\text{par}\right)^2}.
\label{eq3}
\end{equation}

\noindent
If $Y_\text{DUT}$ is a pure capacitor, the minimal detectable change $\delta C_\text{DUT}$ in $C_\text{DUT}$ is therefore given by

\begin{equation}
\delta C_\text{DUT} = \frac{v_\text{noise}}{|A_\text{HEMT} \cdot \omega \cdot S|},
\label{eq4}
\end{equation}

\noindent
where $v_\text{noise}$ is the spectral density of the voltage noise arriving at the input stage of the lock-in amplifier and $\omega$ is $2\pi$ times the measurement frequency.

\begin{figure}[!t]
	\centering
	\includegraphics[width=85mm]{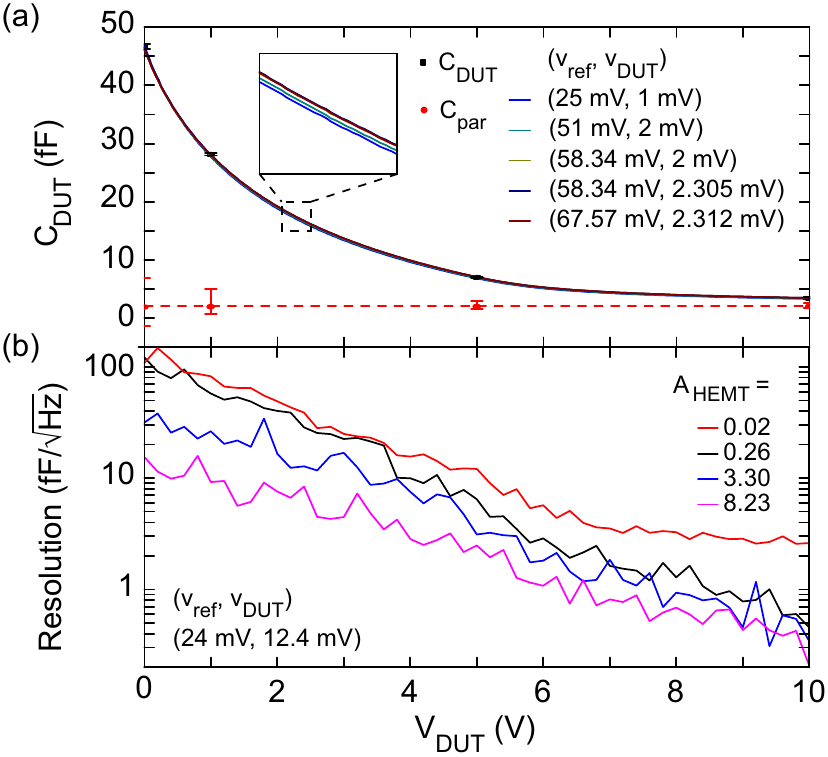}
	\caption{Measurements on a voltage controlled capacitor: (a) the extracted $C_\text{DUT}$ and $C_\text{par}$ from the balance points for different $V_\text{DUT}$ (see text). The continuous lines depict the measurements where we balanced the circuit once using the excitation amplitudes given in the brackets and then swept $V_\text{DUT}$ while monitoring the output $v_{\text out}$. All measurements are in good agreement with one another. (b) the resolution in $C_\text{DUT}$ for different gains of the HEMT while keeping the excitation amplitude given in the brackets fixed. We obtain the best resolution for the highest gain.}
	\label{fig4}
\end{figure}

For the optimization of the resolution to determine $Y_\text{DUT}$, we used a voltage controlled capacitor (SVC704~\cite{onsemiconductor}) as DUT.
These measurements were performed at room temperature.
All measurements in this work were performed at 100~kHz.
The required bias potentials were applied with Yokogawa 7651 DC sources.
The output of the bridge circuit was measured with a Zurich Instruments lock-in amplifier (model UHF).
As we did not observe any phase shifts in $v_\text{out}$, we model $Y_\text{DUT}$ with a capacitance $C_\text{DUT}$.
We balanced the bridge at four different $V_\text{DUT}$ and extracted $C_\text{DUT}$ according to Eq.~(\ref{eq2}) and $C_\text{par}$ from the $v_\text{ref}$ dependence of Eq.~(\ref{eq1}).
The data points in Fig.~\ref{fig4}a show that $C_\text{par}$ is four times bigger than the gate capacitance of the HEMT, which indicates a significant contribution coming from the PCB itself.
We also swept $V_\text{DUT}$ for five different pairs of $v_\text{ref}$ and $v_\text{DUT}$ while recording $v_\text{out}$ and translated this into $C_\text{DUT}$ using Eq.~(\ref{eq1}).
As depicted in Fig.~\ref{fig4}a, the extracted capacitances from all the different measurements are in excellent agreement with one another.
The absolute values are also in good agreement with the datasheet of the voltage controlled capacitor. \cite{onsemiconductor}
Now confident that our bridge circuit gives reproducible data, we balanced the circuit at $V_\text{DUT} = 10$~V and fixed $v_\text{ref}$ and $v_\text{DUT}$.
Then, we measured $v_\text{out}$ as a function of $V_\text{DUT}$ for different HEMT amplifications $A_\text{HEMT}$.
At each $V_\text{DUT}$, we recorded 100~points with a rate of 1~point/s to estimate the noise in $v_\text{out}$ and thus in $C_\text{DUT}$.
We define the resolution as the root-mean-square value of the noise in $C_\text{DUT}$ divided by $\sqrt{BW}$, where $BW$ is the measurement bandwidth.
Figure \ref{fig4}b shows that the resolution in capacitance is best for the highest HEMT amplification and lowest $C_\text{DUT}$.
Both results are understood when considering Eq.~(\ref{eq4}).
The highest HEMT amplification directly minimizes $\delta C_\text{DUT}$ whereas the lowest $C_\text{DUT}$ maximizes $S$ (see Eq.~(\ref{eq3})) and thereby minimizes $\delta C_\text{DUT}$.
The former directly implies that the circuit is not limited by the voltage noise $v_\text{noise}$ arriving at the bridge point, as this would be amplified by the HEMT as well.
We find that the measured $v_\text{noise}$ is approximately equal to the Johnson noise of the 1 M$\Omega$ input resistance of the lock-in amplifier.
Strikingly, the resolution improves by two orders of magnitude when increasing $V_\text{DUT}$ from 0~V to 10~V (see Fig.~3).
As Eq.~(\ref{eq3}) shows, the sensitivity $S$ scales with $C_\text{DUT}^{-2}$, if $|Y_\text{DUT}|$ is (much) larger than $|Y_\text{ref} + Y_\text{par}|$.
Considering this scaling relation and the measured $C_\text{DUT}$ values depicted in Fig.~\ref{fig4}a, the observed improvement in resolution is in agreement with the decrease by one order of magnitude in $C_\text{DUT}$.

\begin{figure}[!t]
	\centering
	\includegraphics[width=85mm]{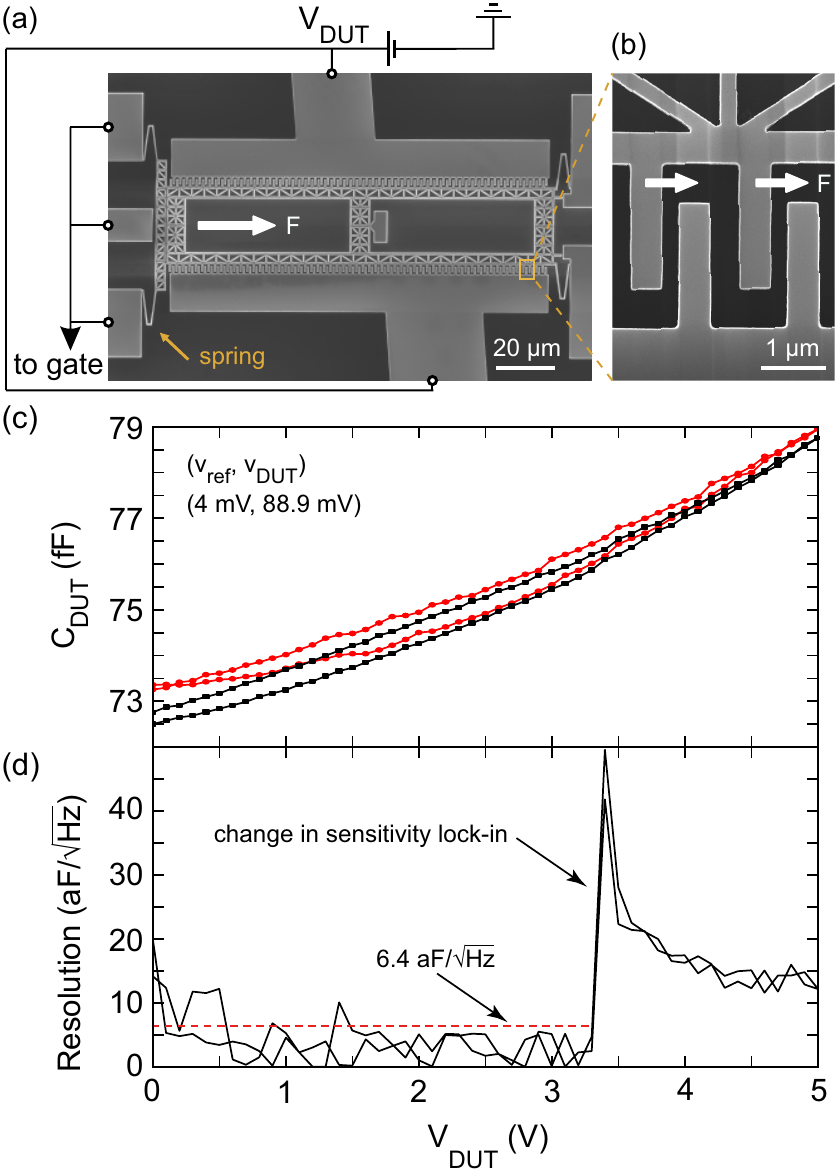}
	\caption{Measurements on a comb-drive actuator: (a) scanning electron microscope of a typical device. The suspended body is held by four springs and is electrically connected to the gate of the HEMT. We apply the voltage $V_\text{DUT}$ to the part that is fixed to the substrate. The zoom in panel (b) shows the asymmetrically placed fingers that give rise to an electrostatic force $F$. (c) the extracted $C_\text{DUT}$ from the balance points for different $V_\text{DUT}$ (red) and from the measurement where we balanced the circuit once using the excitation amplitudes given in the brackets and then swept $V_\text{DUT}$ while monitoring the output $v_\text{out}$ (black). All measurements are in good agreement with one another. (d) the resolution in $C_\text{DUT}$ as a function of $V_\text{DUT}$. The average value for $V_\text{DUT} \leq 3.3$~V is 6.4~aF/$\sqrt{\mathrm{Hz}}$. The sudden change in resolution is due to a change in input sensitivity of the lock-in amplifier.}
	\label{fig5}
\end{figure}

To explore the limits of the achievable resolution with the bridge circuit, we performed measurements on a device with a negligible $|Y_\text{DUT}|$ compared to $|Y_\text{ref} + Y_\text{par}|$ such that the sensitivity $S$ becomes independent from $Y_\text{DUT}$.
For these measurements, we choose a silicon-based micro-machined comb-drive actuator which we fabricate following the process described in Refs. [\onlinecite{goldsche2018raman}] and [\onlinecite{goldsche2018fab}].
In short, the substrate consists from bottom to top of a 500 \textmu m thick undoped Si layer, a 1 \textmu m thick SiO$_\text{2}$ layer, and a 1.2 \textmu m chemical vapor deposited crystalline, highly p-doped silicon layer. The doping of the top layer is $>10^{19}$ cm$^\text{3}$, making our devices low temperature compatible.
Using standard electron beam lithography techniques and reactive ion etching with C$_\text{4}$F$_\text{8}$ and SF$_\text{6}$, we pattern the comb-drive actuators as shown in Fig.~\ref{fig5}a.
The actuator is suspended by etching the SiO$_\text{2}$ underneath the highly p-doped silicon layer away with 10\% HF acid solution.
Finally, a critical-point drying step is used to prevent the comb-drive actuator from collapsing due to capillary forces.
The comb-drive actuator consists of a suspended body that is held by four springs and a part that is fixed to the substrate.
The interdigitated fingers of the suspended body and the fixed part gives rise to an effective parallel plate capacitance of approximately 13~fF.
As the fingers are asymmetrically placed (see Fig.~\ref{fig5}b), the potential $V_\text{DUT}$ applied to the fixed part gives rise to an electrostatic force $F \sim V_\text{DUT}^2$ which results in a displacement of the suspended body and thus into a change of the capacitance.
As the undoped silicon substrate becomes completely insulating around 150~K, we use the impedance bridge at 77~K to measure this capacitance.
This allows us to exclude any effects coming from the undoped silicon.
The output of the bridge circuit was measured with a Stanford Research lock-in amplifier (model SR830).
The measurement was performed with a small $v_\text{\text DUT}$ of 88.9 $\mu$V or 15 $k_\text{B} T/e$ (frequency of 100~kHz).
We did not observe any phase shifts in $v_\text{out}$ and thus model $Y_\text{DUT}$ with only a capacitance $C_\text{DUT}$.
As Fig. \ref{fig5}c shows, the capacitance of this device is 73~fF and increases to 79~fF at $V_\text{DUT} = 5$ V.
We determined $C_\text{par}$ to be 6.5~pF using the same method as for the voltage controlled capacitance discussed above.
The capacitance is a few tens of fF higher than the one from the parallel plate approximation for the interdigitated fingers due to additional capacitances coming from stray fields, bond wires, and other parts of the comb-drive actuator.
The quadratic dependence of $C_\text{DUT}$ on $V_\text{DUT}$, albeit offset such that the minimum $C_\text{DUT}$ is at $V_\text{DUT} \approx -5$ V, is in agreement with the electrostatic force between the interdigitated fingers.
The offset is due to residual charges from the fabrication process. \cite{goldsche2018raman}
The resolution (see Fig. \ref{fig5}d) does, as expected, not depend on $C_\text{DUT}$, which is illustrated by the constant red line below $V_\text{DUT} = 3.3$~V.
The red line illustrates a resolution of 6.4~aF/$\sqrt{\mathrm{Hz}}$, which shows the feasibility of using the bridge circuit for measuring the capacitance of such comb-drive actuators with sub-nm resolution.

\section{Experimental setup and measurements\label{sect2}}

To show the applicability of the impedance bridge at low temperatures in a high magnetic field, we fabricated a hBN/graphene/hBN heterostructure with a gate on top.
We grow the graphene using chemical vapor deposition on a copper foil and then use a dry transfer process to fabricate a hBN/graphene/hBN heterostructure on an undoped Si/SiO$_\text{2}$ substrate. \cite{banszerus2015}
All required hBN sheets are obtained via mechanical exfoliation.
We structure the heterostructures afterwards using electron beam lithography and reactive ion etching to obtain a well defined geometry of $20 \times 20$~$\mathrm{\mu}$m$^2$.
This is subsequently followed by electron beam lithography and Cr/Au evaporation to obtain Ohmic contacts to the graphene sheet.
Then we use the dry transfer process\cite{banszerus2015} to cover the heterostructure and the Ohmic contacts by an additional hBN sheet.
Finally, we use electron beam lithography and Cr/Au evaporation to fabricate a gate on top of the heterostructure.
The area of the top gate is $15 \times 15$~$\mathrm{\mu}$m$^2$ and the hBN layers between the graphene and the top gate are roughly 31~nm thick, which results in a parallel plate capacitance of 0.25~pF when using a relative permittivity of 3.9 for the hBN.
An atomic force microscope image of the final device is shown in Fig.~\ref{fig6}a.
Finally, an Ohmic contact of the device was wirebonded to the impedance bridge.
For completeness, we note that the other two Ohmic contacts were kept floating during the measurement and that the four top gates covering the edge of the graphene were grounded.

Measurements were carried out in a Triton 200 cryostat from Oxford Instruments that is equipped with a superconducting magnet of up to $B = 12$~T.
The HEMT drain voltage $V_\text{dd}$ was biased with 0.55~V to avoid instabilities in its drain-source channel.
The gate of the HEMT was biased with -0.29~V for operation at its highest amplification (see Fig.~\ref{fig3}c).
Due to current flowing through the drain-source channel, the temperature in the cryostat increased from 15~mK to 50~mK.
The impedance bridge is then used to measure the complex admittance from the top gate through the graphene sheet to the gate of the HEMT.
We balance the bridge for a chosen $v_\text{DUT}$ on the order of $k_\text{B}T/e$ by adjusting the amplitude and phase of $v_\text{ref}$.
Note that $v_\text{ref}$ is smaller than $v_\text{DUT}$ for the measurements here.
From the balance point, we extract the complex admittance $Y_\text{DUT}$ using Eq.~(\ref{eq2}).
Then, we sweep the bias $V_\text{DUT}$ of the top gate as a function of an externally applied magnetic field while recording the amplitude and the phase of the output $v_\text{out}$ of the bridge circuit.
We compute the change in $Y_\text{DUT}$ from the deviation of $v_\text{out}$ and thus $v_\text{b}$ away from the balancing point using Eq.~(\ref{eq1}).
As the DC output of the circuit $V_\text{out}$ does not vary while sweeping $V_\text{DUT}$ but the phase of $v_\text{out}$ does (see Figs.~\ref{fig6}b and \ref{fig6}c), we model the complex admittance $Y_\text{DUT}$ by a capacitor and a resistor in series.
We split this capacitance into the geometric capacitance of the top gate to the graphene and the quantum capacitance due to a finite density of states in graphene.
The resistance is attributed to a resistance in the graphene sheet and a contact resistance.

\section{Results and discussion}

\begin{figure*}[!t]
	\centering
	\includegraphics[width=165mm]{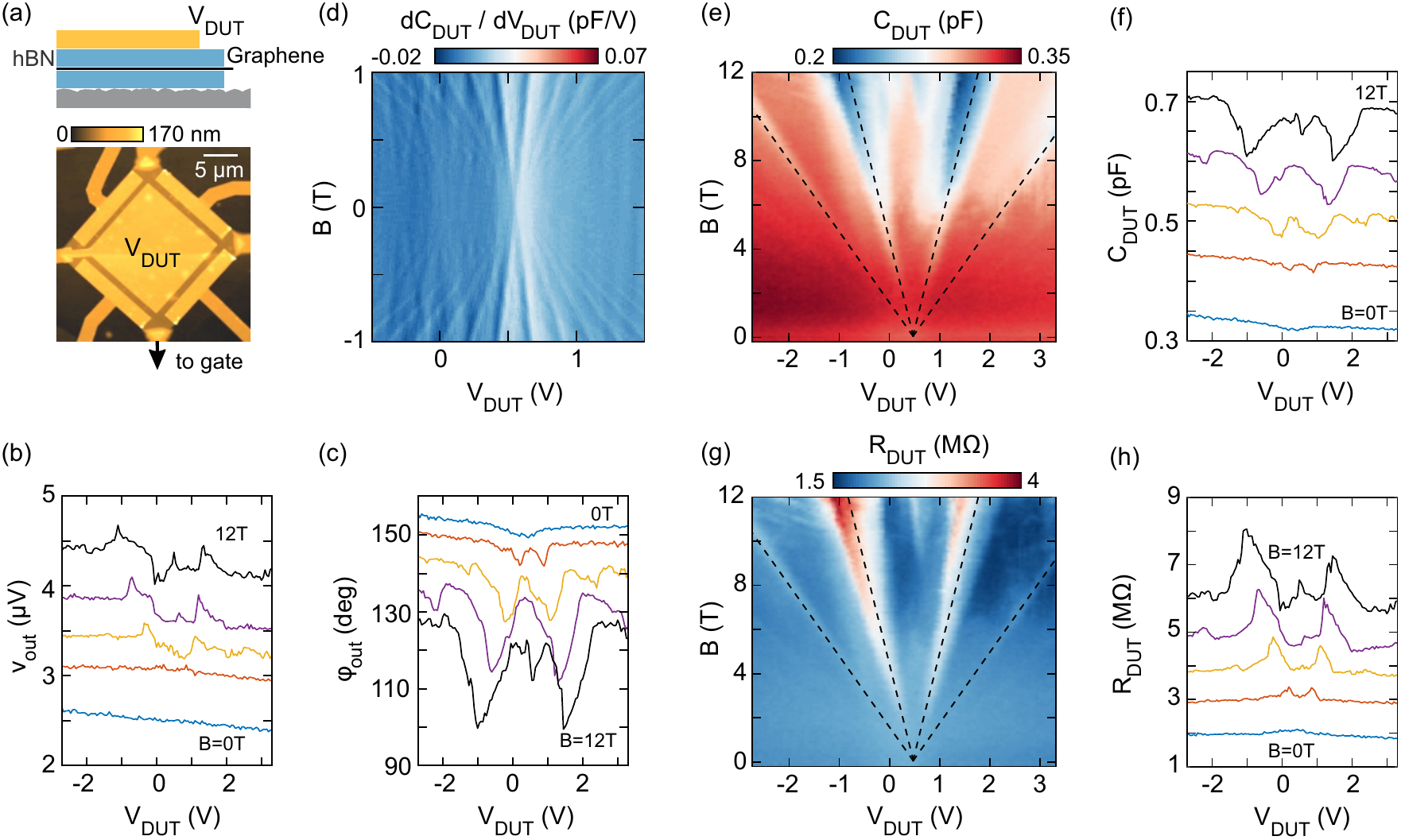}
	\caption{Measurements on a top gated graphene sheet: (a) schematic cross section of the measured device (top panel) and an atomic force microscope image (bottom panel) in which the top gate where $V_\text{DUT}$ is applied and the Ohmic contact that is wirebonded to the gate of the HEMT are indicated. Unprocessed output (b) amplitude and (c) phase of the impedance bridge at magnetic field strengths ranging from 0 to 12~T in steps of 3~T. Note that the curves are offset from another by 1~$\mu$V and 5~degrees for clarity. (d) a high resolution zoom in of (e) the extracted capacitance $C_\text{DUT}$ with line traces in panel (f) for $B$ values ranging from 0 to 12~T in steps of 3~T. (g) the extracted resistance $R_\text{DUT}$ and (h) line traces for $B$ values ranging from 0 to 12~T in steps of 3~T. We observe the formation of Landau levels in both $C_\text{DUT}$ and $R_\text{DUT}$. The slope of the inner pair of black dashed lines is 10.1 T/V and gives us the so-called lever arm of the top gate. The slope of the outer pair of blacked dashed lines is exactly half of it. The line traces in (f) and (h), offset from one another by 0.1~pF and 1~M$\Omega$ for clarity, show that the resistance $R_\text{DUT}$ increases and $C_\text{DUT}$ decreases between Landau levels.}
	\label{fig6}
\end{figure*}

Graphene exhibits an electronic bandstructure, where the conduction and the valence band touch in (two) so-called Dirac points. \cite{katsnelson2012}
The energy-momentum relation is linear around each Dirac point which directly results in a linearly varying density of states and thus a quantum capacitance $C_\text{q}$ that linearly varies with the Fermi energy. \cite{katsnelson2012}
The quantum capacitance has a minimum when the Fermi energy aligns with the Dirac points. \cite{kliros2015}
We control the Fermi energy by the bias $V_\text{DUT}$ applied to the top gate.
When the Fermi energy is far away from the Dirac point, the density of states and thus $C_\text{q}$ will be large such that the geometric capacitance $C_\text{g}$ dominates.
In case of an externally applied magnetic field, discrete Landau levels emerge in the electronic band structure of graphene. \cite{katsnelson2012}
Consequently, the density of states and thus $C_\text{q}$ shows a minimum when the Fermi energy is tuned between two bulk Landau levels.
The resistance $R_\text{DUT}$ is expected to show a maximum when $C_\text{q}$ is minimum due to the absence of states that can contribute to transport. \cite{yu2013,zibrov2017}

Figures~\ref{fig6}d-h show the extracted graphene capacitance and resistance as a function of applied gate bias $V_\text{DUT}$ and magnetic field for our integrated impedance bridge at 60~mK.
The $V_\text{DUT}$ has been shifted by -0.29~V to compensate for the gate bias of the HEMT.
The measurement was performed with a small $v_\text{DUT}$ of 59.3 $\mu$V or 12.9 $k_\text{B} T/e$, which resulted in resolutions of 4.1 k$\Omega/\sqrt{\text Hz}$. and 1.7 fF/$\sqrt{\text Hz}$.
This capacitance resolution is for the given $v_\text{DUT}$ similar to those obtain using an LCR meter. \cite{droscher2010}
We observe the emergence of Landau levels in both the extracted $C_\text{DUT}$ and $R_\text{DUT}$ with increasing applied magnetic field as dips and peaks, i.e. see line traces in Figs.~\ref{fig6}f and \ref{fig6}h.
Even far below 1~T, we can observe the formation of Landau levels (see Fig.~\ref{fig6}d), which illustrates the good quality of the device.
The overall capacitance and resistance curves thus show the trend expected from the electronic band structure of graphene and is in agreement with graphene capacitance and transport measurements reported in the literature. \cite{xia2009,droscher2010,chen2013,yu2013,kliros2015,zibrov2017}
The measured $C_\text{DUT} \sim 0.3$ pF is slightly higher than the expected parallel plate capacitance.
Note that the difference is about equal to that observed for the comb-drive actuator above.
To understand this further, we extracted the slope of the features in $C_\text{DUT}$ and $R_\text{DUT}$ (see black dashed lines in Figs.~\ref{fig6}e and \ref{fig6}g) and we extract the so-called lever arm, which equals the capacitance divided by $e$ and the gate area \cite{Ihn2004}. This gives a capacitance of 0.24~pF.
Interestingly, this value is in agreement with the parallel plate capacitance and it is also lower than the measured capacitance, which suggests the presence of a parasitic capacitance in parallel to the top gate capacitance due to the presence of the other gates.
Note that the observed change in capacitance on the order of 0.1~pF is smaller than the parallel plate capacitance and is, therefore, in agreement with this picture.
The unexpectedly high values of $R_\text{DUT}$ indicate the presence of a large extracted resistance that is in series to the resistance of the top gated area and a contact resistance, which likely originates from the large graphene area not biased by the top gate (see Fig.~\ref{fig6}a).
Note that this extracted resistance may also include contributions from the unbiased graphene area, contacts as well as the wiring.

\section{Conclusion}

We designed and constructed an integrated impedance bridge that operates from room temperature down to 50~mK temperatures.
By placing the HEMT parallel to the externally applied magnetic field, the integrated impedance bridge keeps its functionality in magnetic fields up to at least 12~T.
We find the best resolution when operating the HEMT at the highest gain.
All measurements in this work were performed with excitation amplitudes close to the order of $k_\text{B} T/e$ to ensure minimal heating of the electronic system.
The presented approach enables direct measurements of the capacitance in micro-electromechanical systems such as comb-drive actuators and can thus be used to estimate their displacement or motion.
Using a hBN/graphene/hBN heterostructure, we showed that the presented approach can be used to measure resistance and capacitance
at finite perpendicular magentic fields simultaneously.
The simultaneous measurement of the resistance and the capacitance could simplify the analysis of transport experiments on systems with a low density of states such as 2D materials.

\section{Acknowledgements}

The thank S. Bosco for helpful discussions and proof reading the manuscript.
Support by the ERC (GA-Nr. 280140), the Helmholtz Nano Facility (HNF) \cite{hnf2017} at the Forschungszentrum J\"ulich and the DFG are gratefully acknowledged.
This project has received funding from the European Union’s Horizon 2020
research and innovation program under Grant Agreement No. 785219.
G.V. acknowledges funding by the Excellence Initiative of the German federal and state governments. K.W. and T.T. acknowledge support from the Elemental Strategy Initiative conducted by the MEXT, Japan and the CREST (JPMJCR15F3), JST.

\end{document}